\documentclass[prl,preprint,preprintnumbers,amsmath,amssymb,groupeaddress,superscriptaddress]{revtex4}

\usepackage[pdftex]{graphicx}
\usepackage{dcolumn}
\usepackage{bm}
\usepackage[latin1]{inputenc}
\usepackage{amssymb}   
\usepackage{amsmath}
\usepackage{color}
\usepackage{manfnt}
\usepackage{eucal}

\newcommand{\ud}[1]{{#1^{\dagger}}}

\newcommand{\ket}[1]{\left| #1\right\rangle}

\newcommand{\mean}[1]{\langle#1\rangle}

\begin{document}


\title{Theory of frequency-filtered and time-resolved\\$N$-photon correlations}

\author{E. del Valle} \email{elena.delvalle.reboul@gmail.com}
\affiliation{Physik Department, Technische Universit{\"a}t
  M{\"u}nchen, James-Franck-Str., 85748 Garching, Germany}

\author{A. Gonzalez-Tudela} \affiliation{F{\'i}sica Te{\'o}rica de la
  Materia Condensada, Universidad Aut{\'o}noma de Madrid, 28049,
  Madrid, Spain}

\author{F. P. Laussy} \affiliation{F{\'i}sica Te{\'o}rica de la
  Materia Condensada, Universidad Aut{\'o}noma de Madrid, 28049,
  Madrid, Spain} \affiliation{Walter Schottky Institut, Technische
  Universit{\"a}t M{\"u}nchen, Am Coulombwall 3, 85748 Garching,
  Germany}

\author{C. Tejedor} \affiliation{F{\'i}sica Te{\'o}rica de la Materia
  Condensada, Universidad Aut{\'o}noma de Madrid, 28049, Madrid,
  Spain}

\author{M. J. Hartmann} \affiliation{Physik Department, Technische
  Universit{\"a}t M{\"u}nchen, James-Franck-Str., 85748 Garching,
  Germany}

\date{\today}

\begin{abstract}
  A theory of correlations between $N$ photons of given frequencies
  and detected at given time delays is presented. These correlation
  functions are usually too cumbersome to be computed explicitly.  We
  show that they are obtained exactly through intensity correlations
  between two-level sensors in the limit of their vanishing coupling
  to the system. This allows the computation of correlation functions
  hitherto unreachable. The uncertainties in time and frequency of the
  detection, which are necessary variables to describe the system, are
  intrinsic to the theory. We illustrate the power of our formalism
  with the example of the Jaynes--Cummings model, by showing how
  higher order photon correlations can bring new insights into the
  dynamics of open quantum systems.
\end{abstract}


\maketitle


Photons emerged as a theoretical concept to explain fundamental
properties of the electromagnetic field, such as the relationship
between the energy of light and its frequency, thermal equilibrium of
light and matter or the photo-electric effect. With the advances in
the generation, emission, transmission and detection of photons,
quantum systems are increasingly addressed at the single photon level
and there is a pressing need for generalizations as well as
refinements of the theory of photo-detection~\cite{vogel_book06a}. For
instance, photon correlations combining both their frequency and time
information are now routinely measured in the laboratory. These
experiments have proven extremely powerful in characterising quantum
systems such as a resonantly driven
emitter~\cite{aspect80a,schrama91a,ulhaq12a}, the strong coupling of
light and matter~\cite{press07a,hennessy07a,kaniber08a}, to perform
quantum state tomography~\cite{akopian06a}, to monitor heralded single
photon sources~\cite{moreau01a} or to access spectral diffusion of
single emitters~\cite{sallen10a}.

At this level of fine control of the attributes of the quantum
particles, one needs a theoretical description significantly more
involved than general mathematical statements, such as the
Wiener--Khinchin theorem which assumes abstract and unphysical
properties of the light field. Eberly and W\'odkiewicz, for instance,
have shown how the physics of the detector needs to be included if a
more realistic description of the light field is
required~\cite{eberly77a}. In general, the more detailed is the
characterization of a quantum system, the more necessary it becomes to
describe its measurement.  A bridge between the quantum system and the
observer can be made with the so-called input-output formalism: the
photons \emph{inside} the system, say with operator~$a$ (we consider a
single mode for simplicity), are weakly coupled to an \emph{outside}
continuum of modes, with operators $A_\omega$ (corresponding to their
frequency $\omega$). In the Heisenberg picture, the output field
allows to compute the time-dependent power spectrum of emission as the
density of output photons with frequency~$\omega_1$ at time $T_1$,
i.e., $S^{(1)}(\omega_1,T_1)=\langle
A^\dagger_{\omega_1}(T_1)A_{\omega_1}(T_1)\rangle$. This quantity is
physical only if the uncertainties of detection in both time and
frequency are jointly taken into
account~\cite{eberly77a}. Mathematically, this amounts to adding two
exponential decays in the Fourier transform of the
time-autocorrelation
$S_{\Gamma_1}^{(1)}(\omega_1,T_1)=\frac{\Gamma_1}{2\pi}\iint_{-\infty}^{T_1}dt_1'dt_4'
e^{-\frac{\Gamma_1}{2}(T_1-t_1')}e^{-\frac{\Gamma_1}{2}(T_1-t_4')}
e^{i\omega_1(t_4'-t_1')}\langle\ud{a}(t_1')a(t_4')\rangle$
where~$\Gamma_1$ is interpreted as the linewidth of the detector. This
so-called physical spectrum reduces to the Wiener--Khinchin theorem in
the steady state and in the limit~$\Gamma_1\rightarrow0$.

Extending this result for the detection of two photons was initially
motivated by the Aspect \emph{et al.}  experiment~\cite{aspect80a} of
resonance fluorescence in the Mollow triplet regime~\cite{mollow69a},
where the peaks of the triplet were found to exhibit strong intensity
correlations. These were described theoretically at first by dedicated
methods for the problem at hand, from Cohen--Tannoudji \emph{et al.}
(dressed atom picture)~\cite{cohentannoudji79a,reynaud83a} and
Dalibard \emph{et al.}~(diagrammatic expansion)\cite{dalibard83a}.
The extension of photo-detection in the spirit of Eberly and
W\'odkiewicz by considering two detectors with respective linewidths
$\Gamma_1$ and $\Gamma_2$ was impulsed by Kn\"oll \emph{et
  al.}~\cite{knoll84a} and Arnoldus and
Nienhuis~\cite{arnoldus84a}. The expressions were of general validity,
even though, due to their complexity, the authors still focused on the
particular case of resonance fluorescence for illustration. The
mathematical foundations, shaky in their initial development, were
firmly established in the course of the following
years~\cite{knoll86a,knoll86b,cresser87a}. The multiplicity of photons
requires a careful time ($\mathcal{T}_\pm$) and normal ($:$) ordering
of the operators~\cite{cresser87a,knoll86b}, and it was realized that
it is the time ordering of $\langle{:} A^\dagger_{\omega_1}(T_1)
A_{\omega_1}(T_1) A^\dagger_{\omega_2}(T_2)
A_{\omega_2}(T_2){:}\rangle$ which provides the physical two-photon
spectrum
$S_{\Gamma_1\Gamma_2}^{(2)}(\omega_1,T_1;\omega_2,T_2)=\frac{\Gamma_1\Gamma_2}{(2\pi)^2}
\iint_{-\infty}^{T_1}dt_1'dt_4'e^{-\frac{\Gamma_1}{2}(T_1-t_1')}e^{-\frac{\Gamma_1}{2}(T_1-t_4')}\iint_{-\infty}^{T_2}dt_2'dt_3'
e^{-\frac{\Gamma_2}{2}(T_2-t_2')}e^{-\frac{\Gamma_2}{2}(T_2-t_3')}e^{i\omega_1(t_4'-t_1')}e^{i\omega_2(t_3'-t_2')}\times\langle\mathcal{T}_-[\ud{a}(t_1')\ud{a}(t_2')]\mathcal{T}_+[a(t_3')a(t_4')]\rangle$. Here,
we have defined $\mathcal{T}_+$ (resp. $\mathcal{T}_-$) to order the
operators in a product with the latest time to the far left (resp. far
right)~\cite{vogel_book06a}. Normalising this expression yields the
sought time- and frequency- resolved two-photon correlation function
$g_{\Gamma_1\Gamma_2}^{(2)}(\omega_1,T_1;\omega_2,T_2)=S_{\Gamma_1\Gamma_2}^{(2)}(\omega_1,T_1;\omega_2,T_2)\big/\big[S_{\Gamma_1}^{(1)}(\omega_1,T_1)S_{\Gamma_2}^{(1)}(\omega_2,T_2)\big]$.
It is positive and finite, and reflects that frequency and time of
emission cannot be both measured with arbitrary precision, in
accordance with Heisenberg's uncertainty principle. The limiting
behaviours of $g_{\Gamma_1\Gamma_2}^{(2)}$ defined in this way are
those expected on physical grounds: photons are uncorrelated at
infinite delays,
$\lim_{|T_2-T_1|\rightarrow\infty}g_{\Gamma_1\Gamma_2}^{(2)}(\omega_1,T_1;\omega_2,T_2)=1$~\cite{glauber63b},
and color-blind detectors recover the standard two-time correlators,
$\lim_{\Gamma_1,\Gamma_2\rightarrow\infty}
g_{\Gamma_1\Gamma_2}^{(2)}(\omega_1,T_1;\omega_2,T_2)=g^{(2)}(T_1;T_2)$.
Further generalisation to $N$-photon correlations follows in this way,
adding pairs of operators with their corresponding
integrals~\cite{knoll86a,knoll90a}.

The actual computation of such $g^{(N)}_{\Gamma_1\dots\Gamma_N}$,
however, have proved so far intractable for $N>2$, even for simple
single-mode systems, such as resonance fluorescence or the single mode
laser~\cite{centenoneelen93a}. The case $N=2$ is already demanding and
thus some approximations were made to simplify the
algebra~\cite{nienhuis93a,joosten00a}. More recently, the resonance
fluorescence problem was revisited without approximations but still
for two photons and at zero time delay only~\cite{bel09a}. The main
reason for such limitations is that all the possible time orderings of
the $2N$-time correlator~$\langle\mathcal{T}_-[\ud{a}(t_1')\dots
\ud{a}(t_N')]\mathcal{T}_+[a(t_{N+1}')\dots a(t_{2N}')]\rangle$ result
in $(2N-1)!!2^{N-1}$ independent terms. Furthermore, each of these
correlators requires the application of the quantum regression theorem
$2N-1$ times.  This growth of the complexity makes a direct
computation hopeless for a quantity which is otherwise straightforward
to measure experimentally, merely by detecting photon clicks as
function of time and energy, a technology provided for instance by a
streak camera~\cite{wiersig09a}.

\begin{figure}[t] 
  \centering
  \includegraphics[width=\linewidth]{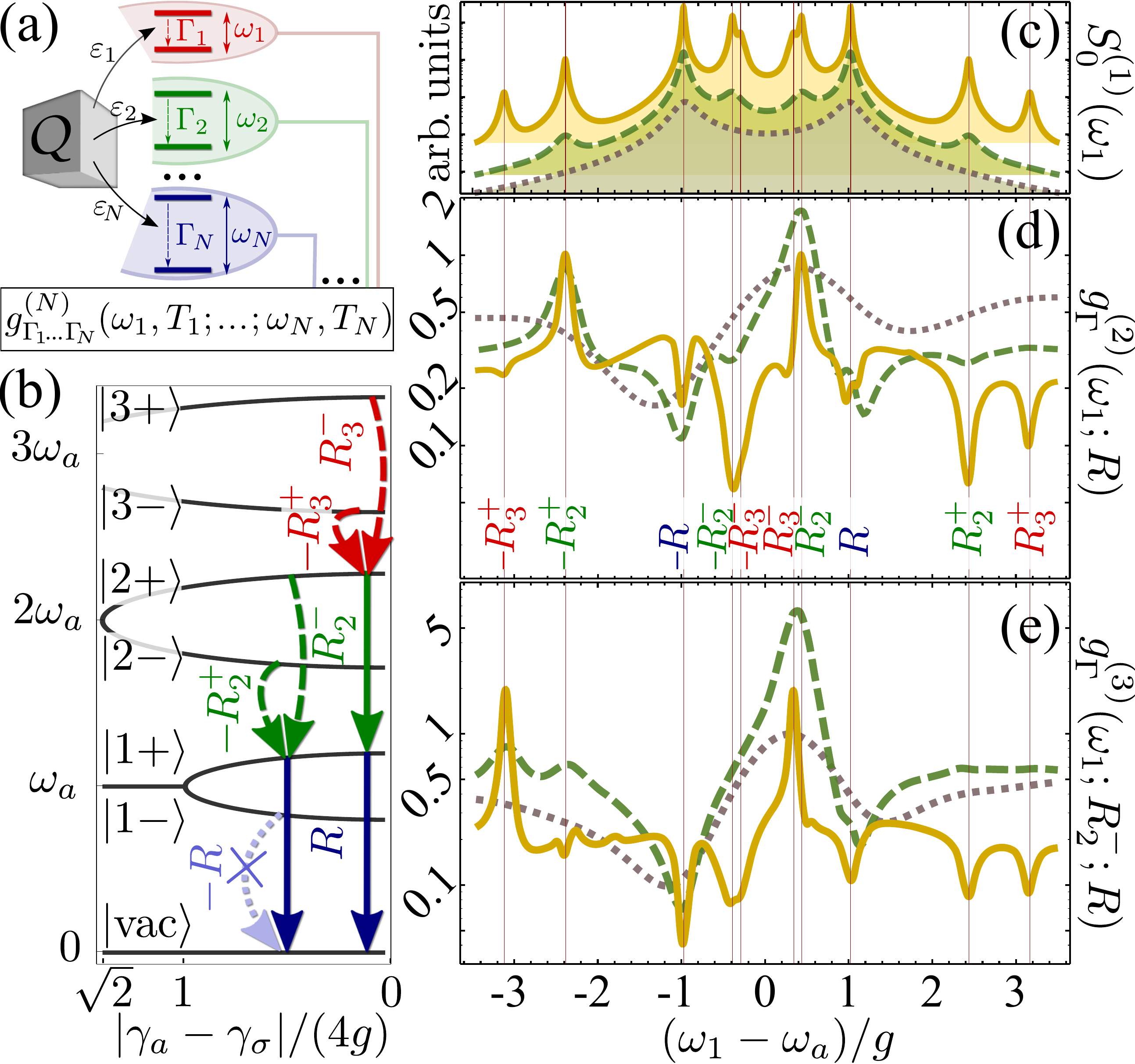}
  \caption{(Color online) (a) Scheme of our proposal to compute
    $N$-photon correlations between photons emitted at different times
    and frequencies from an open quantum open system~$Q$. $N$
    two-level systems of ascribed frequencies are weakly coupled to
    $Q$ and serve as correlation sensors at these frequencies, with
    their decay rate providing the detector linewidth. (b) Dissipative
    Jaynes--Cummings ladder up to the third rung with two of the
    cascades probed in panels (d) [with two sensors] and (e) [with
    three sensors]. Solid arrows show the fixed frequencies. Curved
    arrows show the scanning frequency~$\omega_1$, at the transitions
    where the joint emission is strongly enhanced (dashed) or, on the
    other hand, suppressed (dotted). (c) Power spectra of emission
    probed by weak incoherent excitation
    ($P_\sigma=\gamma_\sigma=0.01g$) for three cavities of decreasing
    quality $\gamma_a=0.01$ (solid), $0.1$ (dashed) and $0.5g$
    (dotted). (d) Two- and (e) three-photon correlations at zero delay
    for the three cavities, with sensor linewidths $\Gamma=\gamma_2$
    (solid) and $\gamma_2/2$ (dashed and dotted).}
  \label{fig:1}
\end{figure}

In this letter, we present a theory of $N$-photon correlations, that
$i)$ allows for arbitrary time delays and frequencies, $ii)$ is
applicable to any open quantum system and $iii)$ is both simple to
implement and powerful. It consists in the introduction of $N$
\emph{sensors} to the dynamics of the open quantum system (noted $Q$
in Fig.~\ref{fig:1}(a)). Each sensor of the set $i=1,\dots,N$ is a
two-level system with annihilation operator $\varsigma_i$ and
transition frequency $\omega_i$, that is matched to the frequency to
be probed in the system. Its lifetime $1/\Gamma_i$ corresponds to the
inverse detector linewidth.  The coupling $\varepsilon_i$ to each
sensor is small enough so that the dynamics of the system is unaltered
by their presence, with
$\mean{n_i}=\mean{\ud{\varsigma_i}\varsigma_i}\ll 1$.  More precisely,
calling $\gamma_Q$ any transition rate within~$Q$ (either with
internal or external degrees of freedom) linked to the field of
interest~$a$, the tunnelling rates $\varepsilon_i$ must be such that
losses into the sensors and their back action are negligible, leading
to $\varepsilon_i\ll\sqrt{\Gamma_i\gamma_Q/2}$.  Under this condition,
we solve the full quantum dynamics of the system supplemented with the
$N$ sensors. The latter play the role of the output fields
$A_{\omega_i}(t)$, but instead of formally solving the Heisenberg
equations and expressing their correlations in terms of the system
operators (as in the standard method exposed above), we compute
directly intensity--intensity correlations between sensors, which is a
considerably simpler task. The main result of this letter, which is
demonstrated in the supplemental material, is:
\begin{equation}
  \label{eq:FriMar16184030CET2012}
  g_{\Gamma_1\dots\Gamma_N}^{(N)}(\omega_1,T_1;\dots ;\omega_N,T_N)=\lim_{\varepsilon_1,\dots,\varepsilon_N\rightarrow 0}\frac{\mean{n_1(T_1)\dots n_N(T_N)}}{\mean{n_1(T_1)}\dots \mean{n_N(T_N)}}
\end{equation}
where the left hand side is the time- and frequency-resolved
$N$-photon correlation function as defined previously~\footnote{Its
  explicit integral form is given for the case $N=2$ in the
  supplemental material, cf.~Eq.~(17) normalized by
  $S_{\Gamma_1}^{(1)}(\omega_1,T_1)S_{\Gamma_2}^{(1)}(\omega_2,T_2)$.}.
The supplemental material establishes that, for open quantum systems
described by Lindblad type master equations, $\langle n_1(T_1)\dots
n_N(T_N)\rangle = \frac{\varepsilon_1^2\dots\varepsilon_N^2}
{\Gamma_1\dots\Gamma_N} (2 \pi)^N
S^{(N)}_{\Gamma_1\dots\Gamma_N}(\omega_1, T_1; \dots; \omega_N, T_N)$
to leading order in the $\varepsilon_i$, which proves
Eq.~(\ref{eq:FriMar16184030CET2012}).  The equality is of general
validity with no approximations or assumptions on the system.  With
this result, the complexity of computing
$g_{\Gamma_1\dots\Gamma_N}^{(N)}(\omega_1,T_1;\dots ;\omega_N,T_N)$ is
greatly reduced as no integral needs to be computed and the quantum
regression theorem needs to be applied $N-1$ times only
. For the important case of zero delay,
$g_{\Gamma_1\dots\Gamma_N}^{(N)}(\omega_1;\dots;\omega_N)$ reduces to
a single-time averaged quantity.  $N$ degenerate sensors with
frequency~$\omega$ and linewidth~$\Gamma$ also provide the $N$th-order
correlations of a single harmonic oscillator with frequency~$\omega$
and linewidth~$\Gamma$, corresponding to the case of correlations
measured after the application of a single filter.  This method is
also useful to derive analytical results (as shown in the supplemental
material).

We now illustrate its efficiency and ease of use by applying it to the
Jaynes--Cummings model~\cite{jaynes63a}, which is both an important
and fundamental quantum description of light-matter
interaction~\cite{shore93a}, is much more complex than resonance
fluorescence as it also quantizes the light field~\cite{delvalle10d}
and is particularly suited to generate strongly correlated
photons~\cite{hartmann08a,reinhard12a}.  Our method recovers exactly
the known results for the Mollow
triplet~\cite{nienhuis93a,joosten00a,bel09a}, and extends them
effortlessly.

At resonance between the light mode ($a$) and the two-level emitter
($\sigma$) both with bare frequency~$\omega_a$, the Jaynes--Cummings
Hamiltonian reads $H=g(\ud{a}\sigma+a\ud{\sigma})$. The master
equation that describes decay ($\gamma_a$, $\gamma_\sigma$) and
incoherent pumping of the emitter ($P_\sigma$) has the form
$\partial_t\rho
=i[\rho,H]+[\frac{\gamma_a}2\mathcal{L}_{a}+\frac{\gamma_\sigma}2\mathcal{L}_{\sigma}+\frac{P_\sigma}2\mathcal{L}_{\ud{\sigma}}](\rho)$,
where $\mathcal{L}_{c}(O)=(2cO\ud{c}-\ud{c}cO-O\ud{c}c)$ and $\rho$ is
the density matrix for the emitter/cavity
system~\cite{delvalle09a}. The new density matrix that includes the
sensors, $\rho_\mathrm{sen}$, follows a modified master equation where
the photonic tunnelling terms,
$H_\mathrm{sen}=\sum_{i=1}^N[\omega_i\ud{\varsigma_i}\varsigma_i+
\varepsilon_i (a \ud{\varsigma_i}+\ud{a}\varsigma_i)]$, are added to
the original Hamiltonian, and the sensor decay terms
$\sum_{i=1}^N\frac{\Gamma_i}2\mathcal{L}_{\varsigma_i}
(\rho_\mathrm{sen})$ are added to the dissipative part.  The level
structure of the dressed states~$\ket{n,\pm}$ with~$n$ excitations is
given by the dissipative Jaynes--Cummings ladder~\cite{delvalle09a},
which is shown in Fig.~\ref{fig:1}(b) at low pumping,
$P_\sigma=\gamma_\sigma$, and in the strong-coupling regime with
$\gamma_\sigma\leq \gamma_a<4g$. This gives rise to the transition
frequencies $R_n^{\pm}=\sqrt{n g^2-\left(
    \frac{\gamma_a-\gamma_\sigma}{4}\right)^2}\pm\sqrt{(n-1)g^2-\left(
    \frac{\gamma_a-\gamma_\sigma}{4}\right)^2}$ between rungs for
$n\ge2$ with broadening
$\gamma_n={2(n-1)\gamma_a+\gamma_\sigma}$~\cite{delvalle09a}. The Rabi
splitting~$2R$, which arises from transitions
$\ket{1\pm}\rightarrow\ket{\mathrm{vac}}$, is given by
$R=\sqrt{g^2-\left(\frac{\gamma_a-\gamma_\sigma}{4}\right)^2}$ with
$\gamma_1=(\gamma_a+\gamma_\sigma)/2$.  These transitions result in
peaks in the power spectrum, as seen in Fig.~\ref{fig:1}(c) for the
three cavity decay rates $\gamma_a/g=0.01$, 0.1 and 0.5 that are
chosen to correspond to cavities embedding superconducting
qubits~\cite{lang11a}, atoms~\cite{koch11a} and quantum
dots~\cite{nomura10a}, respectively. They all show the first rung
transitions at $\pm R$, the so-called Rabi doublet, and one can
distinguish outer peaks at $\pm R_n^{+}$ and inner peaks at $\pm
R_n^{-}$, up to the third rung for the best system (solid line) and to
the second rung for the intermediate one (dashed line).  In
Fig.~\ref{fig:1}(d), we set the linewidth of the sensors~$\Gamma$ at a
value around $\gamma_2$ and compute the two-photon correlation at zero
delay, $g_\Gamma^{(2)}(\omega_1;\omega_2)$, between a photon with
fixed frequency at the Rabi peak, $\omega_2=R$ (solid arrow on the
left of Fig.~\ref{fig:1}(b)), and a photon with variable
frequency~$\omega_1$ which scans the spectral range (curved arrows).
When the scanning frequency $\omega_1$ matches the second rung
transitions that are precursors of the Rabi transition $R$, the
probability of joint emission is enhanced relatively to other
frequencies. The filtering then tracks photons in the cascades
$\ket{2+}\rightarrow\ket{1+}$ at $\omega_1=R_2^{-}$ and
$\ket{2-}\rightarrow\ket{1+}$ at $-R_2^{+}$.  This is a common feature
to all three systems, which shows that even if broadening is too large
to observe explicit features from higher rungs in the power spectrum,
$g^{(2)}_\Gamma(\omega_1;\omega_2)$ allows to uncover them in the
photon correlations. On the other hand, we obtain the expected strong
suppression when the first photon is detected at the other branch of
the Rabi doublet, $\omega_1=-R$. More features can be observed for the
better systems such as dips at the two remaining transitions from the
second rung, $\ket{2-}\rightarrow\ket{1-}$ at $\omega_1=-R_2^{-}$ and
$\ket{2+}\rightarrow\ket{1-}$ at $R_2^{+}$. In the best system, we can
even resolve the dips for the third rung transitions at $\omega_1=\pm
R_3^{\pm}$. All these transitions do not form a consecutive cascade
with the one we fixed and therefore have less probability to occur
within the considered small time window~$1/\gamma_2$.

Instead of making a comprehensive analysis of $g_\Gamma^{(2)}$
specifics, we now turn to higher order correlation functions, such as
the simultaneous three-photon correlations
$g_\Gamma^{(3)}(\omega_1;\omega_2;\omega_3)$, which are exceedingly
hard to compute with previous methods. We fix two frequencies of
detection at $\omega_2=R_2^-$ and $\omega_3=R$ (solid arrows on the
right of Fig.~\ref{fig:1}(b)) and again let $\omega_1$ vary. A strong
enhancement is also observed for all systems, now at
$\omega_1=R_3^{-}$ which monitors the cascade
$\ket{3+}\rightarrow\ket{2+}\rightarrow\ket{1+}\rightarrow\ket{\mathrm{vac}}$
depicted in Fig.~\ref{fig:1}(b) and at $\omega_1=-R_3^+$ which starts
it with $\ket{3-}\rightarrow\ket{2+}$. Other transitions show dips
that are also clearly understood. This hints at the possible
characterization of the level structure of an open quantum system.  In
general, however, one cannot draw conclusions from the zero-delay case
only, in particular for small features, such as the small enhancement
at $\omega_1=-R_2^+$ in $g_\Gamma^{(3)}$ (for the dashed line only)
which is not necessarily a bunching peak and reveals itself in the
$\tau$-dynamics to be antibunched, as discussed later.

\begin{figure}[t] 
  \centering
  \includegraphics[width=\linewidth]{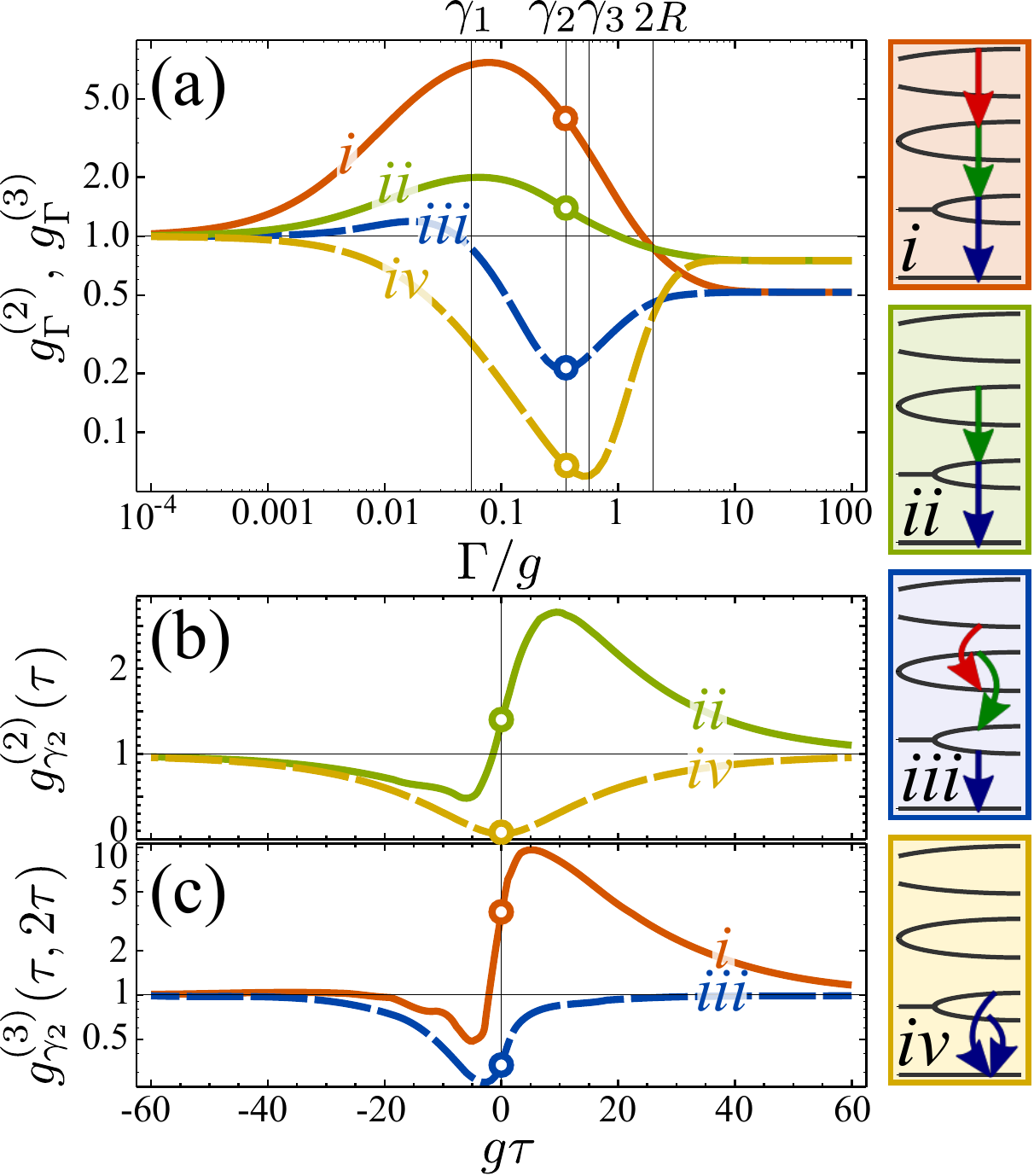}
  \caption{(Color online) (a) Two- and three-photon correlations at
    zero delay as a function of the sensor linewidth $\Gamma$, with
    frequencies of detection as shown in the insets $i$--$iv$.  (b-c)
    $\tau$-dynamics of the correlation functions with
    $\Gamma=\gamma_2$ for, (b), two photons in the configurations of
    insets $ii$ and~$iv$ and, (c), three photons in the
    configurations~$i$ and~$iii$. Positive $\tau$ corresponds to
    detection in the order from top to bottom of the
    ladder. Parameters: $P_\sigma=\gamma_\sigma=0.01g$, $\gamma_a=0.1g$.}
  \label{fig:2}
\end{figure}

In Fig.~\ref{fig:2}(a), we explore another important aspect of
$g_\Gamma^{(N)}$, namely the dependence of correlations on the sensors
linewidths, which is related to the complementary uncertainties in
time and frequency. In the case $\Gamma\rightarrow0$ of perfect
detectors, $g_0^{(N)}=1$ for all~$N$ with non-degenerate frequencies,
since the complete indeterminacy in time leads to averaging photons
from all possible time delays. For $M$ degenerate frequencies out of
$N$, photon indistinguishability results in $M!$~ways for the sensors
to measure the same configuration, that is, $\lim_{\Gamma\rightarrow
  0}g^{(N)}_\Gamma=M!$.  This limit has been misunderstood in the
literature~\footnote{In Ref.~\cite{bel09a}, only the frequency
  convolution is performed and, in the absence of time convolution,
  photon counting diverges in the steady state. A generalized
  Mandel~$\mathrm{Q}$ parameter
  $\sqrt{S_\Gamma^{(1)}(\omega_1)S_\Gamma^{(1)}(\omega_2)}\big(g^{(2)}_\Gamma(\omega_1;\omega_2)-1\big)$
  (in our notations) is used to bypass this difficulty, but for the
  smallest~$\Gamma$ considered, the filtering of the peaks is too
  narrow and the structures obtained are those of the prefactor only
  (uncorrelated photons).}. The effect has otherwise been reported for
the case $M=N=2$ by converting laser light into chaotic light with
narrow filters~\cite{centenoneelen93a}. The other limit
$\Gamma\rightarrow\infty$ corresponds to the opposite situation of
exact $\tau$-delay between photons of completely indeterminate
frequencies. This is of more interest, in particular at zero time
delay, which is the case of Fig.~\ref{fig:2}(a). For the
Jaynes--Cummings system at low pumping, this recovers results derived
by other approaches~\cite{delvalle11a,gartner11a}.

The intermediate case of finite linewidth of the sensors is the most
interesting. Features are the most marked when detector linewidths are
of the order of those of the transitions involved, since the peaks of
the spectrum are best filtered.  Smaller linewidths (longer times) are
to be favoured for bunching and larger linewidths (smaller times) for
antibunching.  One sees for instance in Fig.~\ref{fig:2}(a) that
consecutive transitions, forming a cascade---such as those sketched in
panel $i$ (with three photons) or $ii$ (with two photons)---show an
enhancement. Conversely, the simultaneous emission from both Rabi
peaks, in the configuration sketched as $iv$, is substantially
suppressed, leading to strong antibunching. This observation with a
microcavity containing a single quantum dot has been used to
demonstrate the quantum nature of strong light-matter
coupling~\cite{hennessy07a} (with detuning to better separate the
peaks). Further theoretical investigations with this formalism (to be
discussed elsewhere) may allow to elucidate the nature of spectral
triplets also observed in such
experiments~\cite{hennessy07a,ota09b,gonzaleztudela10b}.

Figures~\ref{fig:2}(b-c) show an example of the $\tau$-dependence of
the correlations, for the case $\Gamma=\gamma_2$, both at positive and
negative delays. The configuration~$ii$ has the typical shape of a
cascade between consecutive levels, with antibunching for $\tau<0$, a
step at $\tau=0$ and bunching for $\tau>0$. This behaviour is well
known, for instance from the biexciton-exciton
cascade~\cite{moreau01a}. It is also observed for $N$ photons in any
consecutive transitions, such as is shown in~$i$ for three photons
starting from the third rung.  In contrast, the filtering of peaks
which do not belong to the same cascade exhibit antibunching, as seen
in $iv$ for the two Rabi peaks or $iii$ for one of its three-photon
counterparts: the order of the transition does not matter anyway and
the cases $\pm\tau$ show qualitatively the same behaviour. These
results are, to the best of our knowledge, the first computations of
three-time frequency-resolved correlation functions.  They are easily
extended to higher orders (a fourth order example is given in the
supplemental material).

In conclusion, we have presented a theory to efficiently compute
correlations between an arbitrary number of photons of any given
frequencies and time delays.  All three aspects of the detection,
namely frequencies, time-delays and linewidths of the detectors, are
needed to characterise meaningfully the system.  The method allows to
compute exactly, with low effort and for general open quantum systems,
properties of output fields that are otherwise defined in terms of
complicated integrals. Its ease of use enabled us to present the first
computation of three and four time-resolved and frequency-filtered
correlation functions.  Its application will allow the interpretation
of experiments which are routinely implemented in the laboratory but
which lacked hitherto an adequate and tractable theoretical support,
and to design new ways to unravel and/or engineer the quantum dynamics
of open systems.

\begin{acknowledgments}
  EdV acknowledges support from the Alexander von Humboldt foundation;
  AGT from the FPU program AP2008-00101 (MICINN); FPL from the Marie
  Curie IEF `SQOD' and the RyC program; CT from MAT2011-22997 (MINECO)
  and S-2009/ESP-1503 (CAM); MJH from the Emmy Noether project HA
  5593/1-1 and from CRC 631 (DFG).
\end{acknowledgments}

\bibliography{Sci,books}

\end{document}



\title{Supplemental Material of ``Theory of Frequency-Filtered and
  Time-Resolved \\ $N$-Photon Correlations''}

\author{E. del
  Valle}\email{elena.delvalle.reboul@gmail.com}\affiliation{Physikdepartment,
  Technische Universit\"at M\"unchen, James-Franck-Stra{\ss}e, 85748
  Garching, Germany}

\author{A. Gonzalez-Tudela} \affiliation{F{\'i}sica Te{\'o}rica de la
  Materia Condensada, Universidad Aut{\'o}noma de Madrid, 28049,
  Madrid, Spain}

\author{F. P. Laussy} \affiliation{Walter Schottky Institut,
  Technische Universit{\"a}t M{\"u}nchen, Am Coulombwall 3, 85748
  Garching, Germany}

\author{C. Tejedor} \affiliation{F{\'i}sica Te{\'o}rica de la Materia
  Condensada, Universidad Aut{\'o}noma de Madrid, 28049, Madrid,
  Spain}

\author{M. J. Hartmann}\affiliation{Physikdepartment, Technische
  Universit\"at M\"unchen, James-Franck-Stra{\ss}e, 85748 Garching,
  Germany}

\date{\today}

\begin{abstract}
  We prove the equality between the experimentally motivated
  correlation function
  $g_{\Gamma_1\ldots\Gamma_N}^{(N)}(\omega_1,T_1;\ldots;\omega_N,T_N)$---defined
  as $2N$-time integrals---and the intensity correlation
  $\mean{n_1(T_1)\ldots
    n_N(T_N)}/(\mean{n_1(T_1)}\ldots\mean{n_N(T_N)})$ between sensors
  coupled to the system, to leading order in the coupling.  We also
  further illustrate the method by calculating correlations in the
  Jaynes--Cummings model up to fourth order.
\end{abstract}

\maketitle

\section{Proof of the equivalence between the sensing and the integral methods}

Let us assume a quantum system described by a set of operators~$a$,
$\sigma$, etc., acting in a Hilbert space~$\mathcal{H}$. In second
quantization, these operators define annihilation operators in the
Heisenberg picture.  The system can be fully Bosonic, Fermionic or a
mixture involving any number of operators. All single-time quantities
can be obtained from correlators of the type
$\mean{a^{\dagger\mu}a^{\nu}\sigma^{\dagger\eta}\sigma^{\theta}\ldots}$
with~$\mu$, $\nu$, $\eta$, $\theta$, etc., integers. Let us
call~$\mathcal{O}$ the set of operators the averages of which
correspond to the correlators required to describe the system, i.e.,
$\mathcal{O}$ includes all the sought observables as well as operators
which couple to them through the equations of motion. In the
following, we assume, without loss of generality, that $a$ is the mode
of interest, the correlations of which are to be computed in time and
frequency.

We prove the case~$N=1$ first, which corresponds to the power
spectrum, then~$N=2$, which corresponds to the most important
correlation function. The proof admits a straightforward
generalization to higher~$N$. The proof proceeds by computing
separately the integral expressions on the one hand and the intensity
correlations between sensors on the other hand, and showing that they
are equal to leading order in the couplings to the sensors. We assume
the steady state case for simplicity, with little loss of generality.

\subsection{$N=1$, power spectrum}

\subsubsection{Integral method}

The single-photon physical spectrum
introduced in the main text as
$S_{\Gamma_1}^{(1)}(\omega_1,T_1)=\frac{\Gamma_1}{2\pi}\times\break\iint_{-\infty}^{T_1}dt_1'dt_4'
e^{-\frac{\Gamma_1}{2}(T_1-t_1')}e^{-\frac{\Gamma_1}{2}(T_1-t_4')}
e^{i\omega_1(t_4'-t_1')}\langle\ud{a}(t_1')a(t_4')\rangle$ can,
through convolutions, be put in the form of an uncertainty in the time
of detection~\cite{eberly77a}:
\begin{equation}
  \label{eq:SatJul14173357CEST2012}
  S_{\Gamma_1}^{(1)}(\omega_1,T_1)=\Gamma_1
  \int_{-\infty}^{T_1} dt_1 e^{-\Gamma_1 (T_1-t_1)}
  \Sigma_{\Gamma_1}^{(1)}(\omega_1,t_1)
\end{equation}
where
\begin{equation}
  \label{eq:SatJul14173450CEST2012}
  \Sigma_{\Gamma_1}^{(1)}(\omega_1,t_1)=\frac{1}{\pi}\Re\int_0^{\infty}
  d\tau_1
  e^{-\frac{\Gamma_1}{2}\tau_1}e^{-i\omega_1\tau_1}\langle\ud{a}(t_1)a(t_1-\tau_1)\rangle
\end{equation}
contains the uncertainty in the frequency of
detection~\cite{nienhuis83a}:
$\Sigma_{\Gamma_1}^{(1)}(\omega_1,t_1)=\int_{-\infty}^{\infty}d\omega_1'
\Sigma_0^{(1)}(\omega_1',t_1)\frac{1}{\pi}\frac{\frac{\Gamma_1}2}{{(\frac{\Gamma_1}{2})^2+(\omega_1'-\omega_1)^2}}$.
The kernel of this expression corresponds to the case of a perfect
detector, $\Gamma_1=0$, known as the Page--Lampard quasi-spectrum of
emission $\Sigma_0^{(1)}(\omega_1,t_1)$~\cite{lampard54a}.  The
results of Eberly and W\'odkiewicz~\cite{eberly77a} show that the
time-dependent physical spectrum~(\ref{eq:SatJul14173357CEST2012}) is
$i)$ always positive, whereas $\Sigma_{0}^{(1)}(\omega_1,t_1)$ is not
in general, and $ii)$ finite, even in the steady state
$S_{\Gamma_1}^{(1)}(\omega_1,T_1\rightarrow\infty)=\Sigma_{\Gamma_1}^{(1)}(\omega_1,t_1)$,
whereas $S^{(1)}(\omega_1,T_1\rightarrow\infty)$ diverges.

To compute Eq.~(\ref{eq:SatJul14173357CEST2012}), we only need to
obtain the two-time correlator
$\langle\ud{a}(t_1)a(t_1-\tau_1)\rangle$. For any two operators $X$
and~$Y$ acting on $\mathcal{H}$, we define the vector
$\mathbf{v}_{X,Y}(\tau)$ as:
\begin{equation}
  \label{eq:ThuMar22124609CET2012}
  \mathbf{v}_{X,Y}(\tau) = \left( \begin{array}{c}
      \mean{X(0)Y(0)} \\
      \mean{X(0)a(\tau)Y(0)} \\
      \mean{X(0)\ud{a}(\tau)Y(0)} \\
      \mean{X(0)(\ud{a}a)(\tau)Y(0)} \\
      \vdots
    \end{array}
  \right)\,,
\end{equation}
where $X$ and~$Y$, in the steady state, sandwich the operators
of~$\mathcal{O}$ taken in some order, which will be kept for the
remainder of the text as starting with the sequence
$\mathcal{O}=\{1, a, \ud{a}, \ud{a}a,\dots\}$.

From the quantum regression theorem, one can define for $\mathcal{O}$
a matrix~$M$ which rules the dynamical evolution of~$\mathbf{v}_{X,Y}$:
\begin{equation}
  \label{eq:TueMar20012758CET2012}
  \partial_\tau \mathbf{v}_{X,Y}(\tau) = M \mathbf{v}_{X,Y}(\tau)\,,
\end{equation}
with solution
$\mathbf{v}_{X,Y}(\tau)=e^{M\tau}\mathbf{v}_{X,Y}(0)$. The steady
state of the system is then fully given by:
\begin{equation}
  \label{eq:TueMar20013537CET2012}  
  \mathbf{v}^\mathrm{ss}=\lim_{\tau\rightarrow
    \infty}\mathbf{v}_{1,1}(\tau)=\lim_{\tau\rightarrow
    \infty}e^{M\tau}\left( \begin{array}{c}
      1 \\
      0 \\
      \vdots
    \end{array}
  \right)\,,
\end{equation}
since $\mathcal{O}$ contains all the relevant observables of the
system. Here we have chosen the vacuum as the initial condition. Since
we employ the standard assumption of a unique steady state, the
initial state does not matter and all the information is encoded in
$e^{M\tau}$.

We now define two matrices, $T_\pm$, which, when acting on
$\mathbf{v}_{X,Y}(\tau)$, introduce an extra $\ud{a}$ for $T_+$ and an
$a$ for $T_-$ between~$X$ and~$Y$, keeping normal ordering:
\begin{equation}
  \label{eq:TueMar20013804CET2012}
  T_+ \mathbf{v}_{X,Y}(\tau)=\left( \begin{array}{c}
      \mean{X(0)\ud{a}(\tau)Y(0)} \\
      \mean{X(0)(\ud{a}a)(\tau)Y(0)} \\
      \mean{X(0)a^{\dagger 2} (\tau)Y(0)} \\
      \mean{X(0)(a^{\dagger 2} a)(\tau)Y(0)} \\
      \vdots
    \end{array}
  \right)\,,
\end{equation}
and 
\begin{equation}
  \label{eq:TueMar20014147CET2012}
 T_- \mathbf{v}_{X,Y}(\tau)=\left( \begin{array}{c}
      \mean{X(0)a(\tau)Y(0)} \\
      \mean{X(0)a^2(\tau)Y(0)} \\
      \mean{X(0)(\ud{a}a)(\tau)Y(0)} \\
      \mean{X(0)(\ud{a} a^2)(\tau)Y(0)} \\
      \vdots
    \end{array}
  \right)\,.
\end{equation}
These matrices always exist, in infinite or in truncated Hilbert
spaces (where, if truncation is to order~$n$, $a^n$ is an operator
in~$\mathcal{O}$ but $a^{n+1}=0$).  For instance, if the mode $a$ is a
two-level system, the vector $\mathbf{v}_{X,Y}(\tau)$ consists of the
first four entries in Eq.~(\ref{eq:ThuMar22124609CET2012}) only, since
$a^{\dagger\mu}a^{\nu}=0$ if $\mu$ or $\nu>1$. Then, these matrices
read
\begin{equation}
T_+ = \left( \begin{array}{lccr}
            0      & 0   & 1 & 0  \\
            0      & 0   & 0 & 1  \\
            0      & 0   & 0 & 0 \\
           0      & 0   & 0 & 0 \\
           \end{array}
    \right)
    \quad\text{and}\quad
T_-= \left( \begin{array}{lccr}
            0      & 1   & 0 & 0  \\
            0      & 0   & 0 & 0  \\
            0      & 0   & 0 & 1 \\
           0      & 0   & 0 & 0 \\
           \end{array}
    \right)\,.
\end{equation}

With these definitions, the correlator
$\langle\ud{a}(t_1)a(t_1-\tau_1)\rangle$ with $\tau_1>0$ is the first
element of $T_+\mathbf{v}_{1,a}(\tau_1)$:
\begin{equation}
  \label{eq:TueMar20023023CET2012}  
  \langle\ud{a}(t_1)a(t_1-\tau_1)\rangle=\left[T_+e^{M\tau_1}T_-\mathbf{v}^\mathrm{ss}\right]_1\,,
\end{equation}
%
where we have used $[\cdots]_i$ to denote the $i$th element of a
vector.  The power spectrum in its integral form is therefore given
by:
\begin{equation}
  \label{eq:ThuApr3171338CEST2008}
  S_{\Gamma_1}^{(1)}(\omega_1)
  =\frac{1}{\pi}\Re\left[T_+\frac{-1}{M+(-i\omega_1-\frac{\Gamma_1}2)\mathbf{1}}T_-\mathbf{v}^\mathrm{ss}\right]_1\,,
\end{equation}
with $\mathbf{1}$ the identity matrix.

\subsection{Sensing method}
\label{sec:sensing1}

We now consider two sensors~$\varsigma_i$, $i=1,2$ with
linewidths~$\Gamma_i$ coupled to the system with
strength~$\varepsilon_i$ such that the dynamics of the system is
probed but is otherwise left unperturbed. This requires the tunnelling
rates $\varepsilon_i$ to fulfil two conditions: the losses into the
sensors must be negligible, $4\varepsilon_i^2/\Gamma_i\ll \gamma_Q$
and so must be the back action of the sensors into the system,
$4\varepsilon_i^2/\gamma_Q\ll \Gamma_i$, where $\gamma_Q$ is the
smallest system decay rate. These conditions both lead to
$\varepsilon_i\ll\sqrt{\Gamma_i\gamma_Q/2}$. We then introduce a
sensing vector~$\mathbf{w}$ of steady state correlators, by
multiplying
$\varsigma_1^{\dagger\mu_1}\varsigma_1^{\nu_1}\varsigma_2^{\dagger\mu_2}\varsigma_2^{\nu_2}$
with the operators in~$\mathcal{O}$:
\begin{equation}
  \label{eq:WedMar21200056CET2012}
  \mathbf{w}[\mu_1\nu_1,\mu_2\nu_2] = \left( \begin{array}{c}
      \mean{\varsigma_1^{\dagger\mu_1}\varsigma_1^{\nu_1}\varsigma_2^{\dagger\mu_2}\varsigma_2^{\nu_2}} \\
      \mean{\varsigma_1^{\dagger\mu_1}\varsigma_1^{\nu_1}\varsigma_2^{\dagger\mu_2}\varsigma_2^{\nu_2}a} \\
      \mean{\varsigma_1^{\dagger\mu_1}\varsigma_1^{\nu_1}\varsigma_2^{\dagger\mu_2}\varsigma_2^{\nu_2}\ud{a}} \\
      \mean{\varsigma_1^{\dagger\mu_1}\varsigma_1^{\nu_1}\varsigma_2^{\dagger\mu_2}\varsigma_2^{\nu_2}\ud{a}a} \\
      \vdots
    \end{array}
  \right)\,,
\end{equation}
where the indices $\mu_i$ and $\nu_i$ take the values 0 or 1. In the
regime under consideration, the population
$\langle\ud{\varsigma_i}\varsigma_i\rangle\ll1$ and the equations of
motion are valid to leading order in~$\varepsilon_{1,2}$:
\begin{widetext}
\begin{multline}
  \label{eq:FriMar30165907CEST2012}
  \partial_t \mathbf{w}[\mu_1\nu_1,\mu_2\nu_2]= \large\{ M+[(\mu_1-\nu_1)i\omega_1-(\mu_1+\nu_1){\frac{\Gamma_1}2}
  +(\mu_2-\nu_2)i\omega_2-(\mu_2+\nu_2)\frac{\Gamma_2}2]\mathbf{1} \large\}
  \mathbf{w}[\mu_1\nu_1,\mu_2\nu_2]\\
  +\mu_1(i\varepsilon_1T_+)\mathbf{w}[0\nu_1,\mu_2\nu_2]+\nu_1(-i\varepsilon_1T_-)\mathbf{w}[\mu_10,\mu_2\nu_2]
  +\mu_2(i\varepsilon_2T_+)\mathbf{w}[\mu_1\nu_1,0\nu_2]+\nu_2(-i\varepsilon_2T_-)\mathbf{w}[\mu_1\nu_1,\mu_20]\,,
\end{multline}
and can be solved recursively:
  \begin{multline}
    \label{eq:FriMar30202508CEST2012}
    \mathbf{w}[\mu_1\nu_1,\mu_2\nu_2]=\frac{-1}{M+[(\mu_1-\nu_1)i\omega_1-(\mu_1+\nu_1)\frac{\Gamma_1}2+(\mu_2-\nu_2)i\omega_2-(\mu_2+\nu_2)\frac{\Gamma_2}2]\mathbf{1}}\\
    \times \Big\{
    \mu_1(i\varepsilon_1T_+)\mathbf{w}[0\nu_1,\mu_2\nu_2]+\nu_1(-i\varepsilon_1T_-)\mathbf{w}[\mu_10,\mu_2\nu_2]
    +\mu_2(i\varepsilon_2T_+)\mathbf{w}[\mu_1\nu_1,0\nu_2]+\nu_2(-i\varepsilon_2T_-)\mathbf{w}[\mu_1\nu_1,\mu_20]\Big\}
    \,.
  \end{multline}
Higher order terms will cancel exactly in the vanishing coupling we
will assume later and thus do not need to be included here. Besides,
unlike the leading order term, the higher order ones depend on the
modelling of the sensors (as two-level systems, harmonic oscillators,
etc.) and on the system itself.
\end{widetext}

The spectrum of emission of $a$ is given by the average population, in
the steady state, of any one of the two sensors, say, the first one:
$\mean{n_1}=\mean{\varsigma^\dagger_1\varsigma_1}$. Its equation of
motion reads $\partial_t \mean{n_1} =-\Gamma_1\mean{n_1}+2\Re
(i\varepsilon_1 \mean{\varsigma_1 \ud{a}})$, and with the above
notations, is therefore given in the steady state by:
\begin{equation}
  \label{eq:WedMar21032339CET2012}
  \mean{n_1}=\frac{2}{\Gamma_1}\Re\Big[i\varepsilon_1T_+ \mathbf{w}[01,00]\Big]_1\,.
\end{equation}
Using the solution Eq.~(\ref{eq:FriMar30202508CEST2012}), the
correlator of interest for the spectrum reads:
\begin{equation}
  \label{eq:WedMar21033037CET2012}
  \mathbf{w}[01,0,0]=\frac{-1}{M+[-i\omega_1-\frac{\Gamma_1}{2}]\mathbf{1}}(-i\varepsilon_1T_-)\mathbf{v}^\mathrm{ss}\,.
\end{equation}

\subsubsection{Equality of the integral and sensing methods}

The proof is now complete since, to leading order, we find that
Eq.~(\ref{eq:ThuApr3171338CEST2008}) and
Eqs.~(\ref{eq:WedMar21032339CET2012}-\ref{eq:WedMar21033037CET2012})
provide the claimed identity:
\begin{multline}
  \label{eq:WedMar21033437CET2012}
  \mean{n_1}=\frac{2\varepsilon_1^2}{\Gamma_1}\Re \left[T_+  \frac{-1}{M+[-i\omega_1-\frac{\Gamma_1}{2}]\mathbf{1}} T_-\mathbf{v}^\mathrm{ss}\right]_1\\
  =\frac{\varepsilon_1^2}{\Gamma_1}(2\pi) S_{\Gamma_1}^{(1)}(\omega_1)\,.
\end{multline}

\subsection{$N=2$, two-photon correlations}

\subsubsection{Integral method}

The case $N=2$ brings with the multiplicity of photons the conceptual
difficulty of time- and normal-ordering. It was discussed in the text
that the proper definition yielding a physical two-photon spectrum
reads~\cite{knoll86b,cresser87a}:
\begin{widetext}
\begin{multline}
  \label{eq:SatJul14180625CEST2012}
  S_{\Gamma_1\Gamma_2}^{(2)}(\omega_1,T_1;\omega_2,T_2)=\frac{\Gamma_1\Gamma_2}{(2\pi)^2}
  \iint_{-\infty}^{T_1}dt_1'dt_4'e^{-\frac{\Gamma_1}{2}(T_1-t_1')}e^{-\frac{\Gamma_1}{2}(T_1-t_4')}\iint_{-\infty}^{T_2}dt_2'dt_3'e^{-\frac{\Gamma_2}{2}(T_2-t_2')}e^{-\frac{\Gamma_2}{2}(T_2-t_3')}\\
  \times e^{i\omega_1(t_4'-t_1')}e^{i\omega_2(t_3'-t_2')}\langle\mathcal{T}_-[\ud{a}(t_1')\ud{a}(t_2')]\mathcal{T}_+[a(t_3')a(t_4')]\rangle\,.
\end{multline}
In analogy with the case $N=1$, it can be put in the form:
\begin{equation}
  \label{eq:FriMar16153030CET2012}
  S_{\Gamma_1\Gamma_2}^{(2)}(\omega_1,T_1;\omega_2,T_2)=
  {\Gamma_1\Gamma_2}\int_{-\infty}^{T_1}dt_1\int_{-\infty}^{T_2}dt_2
  e^{-\Gamma_1(T_1-t_1)}e^{-\Gamma_2(T_2-t_2)}\Sigma_{\Gamma_1\Gamma_2}^{(2)}(\omega_1,t_1;\omega_2,t_2)\,,
\end{equation}
isolating the \emph{two-photon quasi-distribution}:
\begin{multline}
  \label{eq:FriMar16154500CET2012}
  \Sigma_{\Gamma_1\Gamma_2}^{(2)}(\omega_1,t_1;\omega_2,t_2)=\frac{2\Re}{(2\pi)^2}\iint_{0}^{\infty} d\tau_1d\tau_2 e^{-\frac{\Gamma_1}{2}\tau_1}e^{-\frac{\Gamma_2}{2}\tau_2} \\
  \times e^{-i\omega_2\tau_2}[e^{i\omega_1\tau_1}\langle\mathcal{T}_-[\ud{a}(t_1-\tau_1)\ud{a}(t_2)]\mathcal{T}_+[a(t_2-\tau_2)a(t_1)]\rangle
  +e^{-i\omega_1\tau_1}\langle\mathcal{T}_-[\ud{a}(t_1)\ud{a}(t_2)]\mathcal{T}_+[a(t_2-\tau_2)a(t_1-\tau_1)]\rangle]\,,
\end{multline}
which, like the quasi-spectrum, can be negative and is thus not a
physical spectrum.

\end{widetext}
To proceed with the calculation, let us separate the $\tau=T_2-T_1$
two-photon correlation function between its $\tau=0$ and $\tau>0$
terms:
\begin{multline}
  \label{eq:FriMar16153030CET2012}
  S_{\Gamma_1\Gamma_2}^{(2)}(\omega_1;\omega_2,\tau)=\\
  e^{-\Gamma_2\tau}S_{\Gamma_1\Gamma_2}^{(2)}(\omega_1;\omega_2)+\Delta
    S_{\Gamma_1\Gamma_2}^{(2)}(\omega_1;\omega_2,\tau)\,,
\end{multline}
with
\begin{multline}
  \label{eq:FriMar16153949CET2012}
  S_{\Gamma_1\Gamma_2}^{(2)}(\omega_1;\omega_2)=\Gamma_1\Gamma_2\int_{-\infty}^{T_1}dt_2 \int_{-\infty}^{t_2} dt_1\\
  \times e^{-\Gamma_1(T_1-t_1)}e^{-\Gamma_2(T_1-t_2)}\Sigma_{\Gamma_1\Gamma_2}^{(2)}(\omega_1,t_1;\omega_2,t_2)+\left[ 1\leftrightarrow 2\right]\,,
\end{multline}
and
\begin{multline}
  \label{eq:FriMar30194947CEST2012}
  \Delta S_{\Gamma_1\Gamma_2}^{(2)}(\omega_1;\omega_2,\tau)=\Gamma_1\Gamma_2\int_{T_1}^{T_2}dt_2\int_{-\infty}^{T_1}dt_1\\
  \times
  e^{-\Gamma_1(T_1-t_1)}e^{-\Gamma_2(T_2-t_2)}\Sigma_{\Gamma_1\Gamma_2}^{(2)}(\omega_1,t_1;\omega_2,t_2)\,,
\end{multline}
where $\left[ 1\leftrightarrow 2\right]$ means the interchange of
sensors 1 and 2, that is, permuting $\omega_1\leftrightarrow \omega_2$
and $\Gamma_1\leftrightarrow \Gamma_2$.

To compute these quantities, it is enough to consider
$\Sigma_{\Gamma_1\Gamma_2}^{(2)}(\omega_1,t_1;\omega_2,t_2)$ for
$t=t_2-t_1>0$ since the inverse order is given by the exchange $\left[
  1\leftrightarrow 2\right]$. Therefore, we restrict the integration
to ordering of the time variables where $t_1-\tau_1<t_1<t_2$. The
fourth variable yields three different domains of integration:
\begin{itemize}
\addtolength{\itemsep}{-\baselineskip}
\item[(1)] $t_2-\tau_2<t_1-\tau_1<t_1<t_2$, 
\item[(2)] $t_1-\tau_1<t_2-\tau_2<t_1<t_2$,
\item[(3)] $t_1-\tau_1<t_1<t_2-\tau_2<t_2$.
\end{itemize}
For each of them, there are two different correlators appearing in
$\Sigma^{(2)}$: one with the factor
$e^{-i\omega_2\tau_2}e^{i\omega_1\tau_1}$, the other with
$e^{-i\omega_2\tau_2}e^{-i\omega_1\tau_1}$. They will be respectively
referred to as $\mathcal{C}_{(ia)}$ and $\mathcal{C}_{(ib)}$, with
$i=1,2,3$ depending on their domains of integration. This gives rise
to six integrals which we shall denote $\mathcal{I}_{(ia)}$ and
$\mathcal{I}_{(ib)}$.

From this discussion, we can find a general expression for the
complexity of the integration method in terms of the various domains
of integration and the different correlators to be considered. The
number of independent time ordering is $(2N-1)!!$ and the number of
independent terms in $\Sigma^{(N)}$ is $2^{N}/2$ (we divide by 2
because half are complex conjugates of the other half). The total
number of independent time integrals and correlators is therefore
$(2N-1)!!2^{N-1}$.

The first correlator we need,
$\mathcal{C}_{(1a)}=\mean{\ud{a}(t_1-\tau_1)\ud{a}(t_2)a(t_1)a(t_2-\tau_2)}$,
is the first element of the vector $T_+\mathbf{v}_{X_1,Y_1}(t)$ with
$X_1=\ud{a}(t_1-\tau_1)$ and $Y_1=a(t_1)a(t_2-\tau_2)$. We obtain
$\mathbf{v}_{X_1,Y_1}(t)= e^{M t}\mathbf{v}_{X_1,Y_1}(0)=
e^{Mt}T_-\mathbf{v}_{X_1,Y_2}(\tau_1)$ with $Y_2=a(t_2-\tau_2)$. In
turn,
$\mathbf{v}_{X_1,Y_2}(\tau_1)=e^{M\tau_1}\mathbf{v}_{X_1,Y_2}(0)=e^{M\tau_1}T_+\mathbf{v}_{1,Y_2}(t')$
is obtained with $Y_2=a(t_2-\tau_2)$ and
$t'=\tau_2-\tau_1-t$. Finally, we get
$\mathbf{v}_{1,Y_2}(t')=e^{Mt'}\mathbf{v}_{1,Y_2}(0)=e^{Mt'}T_-
\mathbf{v}^\mathrm{ss}$. Putting everything together, we get:
\begin{equation}
  \label{eq:WedMar21042522CET2012}
  \mathcal{C}_{(1a)\atop(1b)}= \left[T_+e^{Mt}T_\mp e^{M\tau_1}T_\pm e^{Mt'}T_- \mathbf{v}^\mathrm{ss}\right]_1\,,
\end{equation}
with correspondence between upper and lower indices with the
sign. Repeating this procedure for the other domains of integration,
we also get:
  \begin{equation}
  \label{eq:WedMar21181637CET2012}
    \mathcal{C}_{(2a)\atop(2b)}= \left[T_+e^{Mt}T_\mp e^{-Mt}e^{M \tau_2}T_- e^{Mt''}T_\pm\mathbf{v}^\mathrm{ss}\right]_1\,,
\end{equation}
where we defined $t''=t+\tau_1-\tau_2$ (going from $0$ to
$\infty$), and
\begin{equation}
  \label{eq:WedMar21182123CET2012}
  \mathcal{C}_{(3a)\atop(3b)}= \left[T_+e^{M\tau_2}T_-e^{Mt}e^{-M \tau_2}T_\mp e^{M\tau_1}T_\pm\mathbf{v}^\mathrm{ss}\right]_1\,.
\end{equation}

\subsubsection{Integral method at $\tau=0$}

We now turn to the zero time delay contribution
$S_{\Gamma_1\Gamma_2}^{(2)}(\omega_1;\omega_2)$, which, according to
Eq.~(\ref{eq:FriMar16153949CET2012}), is given by integrating the
correlators~(Eqs.~(\ref{eq:WedMar21042522CET2012}--\ref{eq:WedMar21182123CET2012}))
over their corresponding domains, changing variables as needed. For
instance, the integrals of correlators $\mathcal{C}_{(1i)}$ require
the change of variables $t_1\rightarrow t$ and $\tau_2 \rightarrow t'$
(both extending from $0$ to $\infty$). The final expressions for the
two integrals $(a)$ and $(b)$ read:
\begin{multline}
  \label{eq:MonMar26124522CEST2012}
  \mathcal{I}_{(1a)\atop(1b)}= \frac{\Gamma_1\Gamma_2}{\Gamma_1+\Gamma_2}\frac{1}{(2\pi)^2}\Big[T_+\frac{-1}{M+(-i\omega_2-\Gamma_1-\frac{\Gamma_2}2)\mathbf{1}}\\
    \times T_\mp \frac{-1}{M+(\pm i\omega_1-i\omega_2-\frac{\Gamma_1+\Gamma_2}2)\mathbf{1}}\\
    \times T_\pm \frac{-1}{M+(-i\omega_2-\frac{\Gamma_2}2)\mathbf{1}}T_- \mathbf{v}^\mathrm{ss}\Big]_1\,.
\end{multline}
The second correlators $\mathcal{C}_{(2i)}$ lead to:
\begin{multline}
  \label{eq:MonMar26124508CEST2012}
  \mathcal{I}_{(2a)\atop(2b)}= \frac{\Gamma_1\Gamma_2}{\Gamma_1+\Gamma_2}\frac{1}{(2\pi)^2}\Big[T_+\frac{-1}{M+(-i\omega_2-\Gamma_1-\frac{\Gamma_2}2)\mathbf{1}}\\
    \times T_\mp \frac{-1}{M+(\pm i\omega_1-i\omega_2-\frac{\Gamma_1+\Gamma_2}2)\mathbf{1}}\\
    \times T_- \frac{-1}{M+(\pm i\omega_1-\frac{\Gamma_1}2)\mathbf{1}}T_\pm \mathbf{v}^\mathrm{ss}\Big]_1\,.
\end{multline}
And the third correlators $\mathcal{C}_{(3i)}$ lead to:
\begin{multline}
  \label{eq:MonMar26124440CEST2012}
  \mathcal{I}_{(3a)\atop(3b)}= \frac{\Gamma_1\Gamma_2}{\Gamma_1+\Gamma_2}\frac{1}{(2\pi)^2}\Big[T_+\frac{-1}{M+(-i\omega_2-\Gamma_1-\frac{\Gamma_2}2)\mathbf{1}}\\
  \times T_- \frac{-1}{M-\Gamma_1\mathbf{1}}\\
  \times T_\mp \frac{-1}{M+(\pm
    i\omega_1-\frac{\Gamma_1}2)\mathbf{1}}T_\pm
  \mathbf{v}^\mathrm{ss}\Big]_1\,.
\end{multline}

The total correlation function follows from twice the real part of the
six previous integrals summed over and exchanging photons:
\begin{equation}
  \label{eq:MonMar26124543CEST2012}
  S_{\Gamma_1\Gamma_2}^{(2)}(\omega_1;\omega_2)= 2\Re \sum_{i=1,2,3}\Big[ \mathcal{I}_{(ia)}+\mathcal{I}_{(ib)} \Big] +\left[ 1\leftrightarrow 2\right]\,.
\end{equation}

\subsubsection{Integral method at $\tau>0$}

The finite time-delay contribution $\Delta
S_{\Gamma_1\Gamma_2}^{(2)}(\omega_1;\omega_2,\tau)$ requires
different domains of integration only for the variables $t_2$, now
ranging from $T_1$ to $T_2$, and $t=t_2-t_1$ now ranging from
$t_2-T_1$ to $\infty$. As a result, the integrals in
Eq.~(\ref{eq:FriMar30194947CEST2012}) depend on $\tau$. The integrals
on the correlators $\mathcal{C}_{(1a)\atop(1b)}$ and
$\mathcal{C}_{(2a)\atop(2b)}$, that we note $\Delta
\mathcal{I}_{(1a)\atop(1b)}$ and $\Delta \mathcal{I}_{(2a)\atop(2b)}$
give similar results as the corresponding
$\mathcal{I}_{(ia)\atop(ib)}$, but they acquire the $\tau$-dependence
in the form of a factor $(\Gamma_1+\Gamma_2)\mathcal{F}(\tau)$, with
\begin{equation}
  \label{eq:ThuMar29193657CEST2012}
  \mathcal{F}(\tau)=e^{-\Gamma_2\tau}\frac{e^{[M+(-i\omega_2+\frac{\Gamma_2}2)\mathbf{1}]\tau}
    -1}{M+(-i\omega_2+\frac{\Gamma_2}2)\mathbf{1}}\,,
\end{equation}
that is to be inserted in Eqs.~(\ref{eq:MonMar26124522CEST2012}),
(\ref{eq:MonMar26124508CEST2012}) after the first matrix $T_+$. The
integrals on $\mathcal{C}_{(3a)\atop(3b)}$, on the other hand, are not
so straightforward. They are to be separated into two parts: one where
$t_2-\tau_2<T_1$, the other one $t_2-\tau_2>T_1$. The first part, with
integrals
$\int_{T_1}^{T_2}dt_2\int_{\tau_2}^{\infty}dt\int_{t_2-T_1}^{\infty}d\tau_2\int_0^{\infty}d\tau_1(\ldots)$,
gives rise to a quantity similar to $\Delta
\mathcal{I}_{(ia)\atop(ib)}(\tau)$ with $i=1,2$, in that its
$\tau$-dependence also consists in the factor
$(\Gamma_1+\Gamma_2)\mathcal{F}(\tau)$ inserted after the first matrix
$T_+$ in Eq.~(\ref{eq:MonMar26124440CEST2012}). For this reason we
note it $\Delta \mathcal{I}_{(3a)\atop(3b)} (\tau)$. The second part,
with integrals
$\int_{T_1}^{T_2}dt_2\int_{t_2-T_1}^{\infty}dt\int_{0}^{t_2-T_1}d\tau_2\int_0^{\infty}d\tau_1(\ldots)$,
yields two more contributions:
\begin{multline}
  \label{eq:ThuMar29201440CEST2012}
  \Delta \mathcal{I}_{(3\alpha)\atop(3\beta)}(\tau)=\frac{\Gamma_1\Gamma_2}{(2\pi)^2}\Big[T_+ \mathcal{Z}(\tau)\\
  \times \frac{-1}{M-\Gamma_1\mathbf{1}} T_\mp \frac{-1}{M+(\pm
    i\omega_1-\frac{\Gamma_1}2)\mathbf{1}}T_\pm
  \mathbf{v}^\mathrm{ss}\Big]_1\,,
\end{multline}
where we introduced the $\tau$-dependent matrix
\begin{multline}
  \mathcal{Z}(\tau)=\int_{T_1}^{T_2}dt_2\int_{0}^{t_2-T_1}d\tau_2e^{-\Gamma_2(T_2-t_2)}\\
  \times e^{\big(M+(-i\omega_2-\frac{\Gamma_2}2)\mathbf{1}\big)\tau_2}T_-e^{M(t_2-T_1-\tau_2)}\,.
\end{multline}
$\mathcal{Z}(\tau)$ can be calculated for each element $T_-^{kl}\in\{0,1\}$ of the
matrix $T_-$:
\begin{multline}
  \label{eq:ThuMar29194113CEST2012}
  \mathcal{Z}_{i,j}(\tau)=e^{-\Gamma_2\tau}\sum_{p,k,l,q}\frac{E_{ip}E^{-1}_{pk}T_-^{kl}E_{lq}E^{-1}_{qj}}{m_p-m_q-i\omega_2-\frac{\Gamma_2}2}\\
  \times \Big\{\frac{e^{(m_p-i\omega_2+\frac{\Gamma_2}2)\tau}-1}{m_p-i\omega_2+\frac{\Gamma_2}2}-\frac{e^{(m_q+\Gamma_2)\tau}-1}{m_q+\Gamma_2} \Big\}\,,
\end{multline}
where $E$ is the matrix of eigenvectors of $M$, that diagonalises it:
$M_{ik}=\sum_{p}E_{ip} m_p E^{-1}_{pk}$, with $m_p$ the eigenvalues.

Gathering terms with the same $\tau$ dependence defines
$\Delta\mathcal{I}(\tau)=\sum_{i=1,2,3}\Big[\Delta \mathcal{I}_{(ia)}
(\tau)+\Delta\mathcal{I}_{(ib)}(\tau) \Big]$ which enters in the final
result:
%
\begin{equation}
  \label{eq:ThuMar29152220CEST2012}
  \Delta S_{\Gamma_1\Gamma_2}^{(2)}(\omega_1;\omega_2,\tau) = 2\Re\Big[ \Delta \mathcal{I}(\tau)+ \Delta \mathcal{I}_{(3\alpha)}(\tau)+ \Delta \mathcal{I}_{(3\beta)}(\tau)\Big]\,.
\end{equation}

\subsubsection{Sensing method at $\tau=0$}

The intensity correlations between two sensors,
$\mean{n_1n_2}=\mean{\varsigma^\dagger_1\varsigma_1\varsigma^\dagger_2\varsigma_2}$,
have the equation of motion:
\begin{multline}
  \partial_t \mean{n_1n_2} =-(\Gamma_1+\Gamma_2)\mean{n_1n_2}
  \\+2\Re \Big[ i\varepsilon_2 \mean{\varsigma^\dagger_1\varsigma_1\varsigma_2
    \ud{a}}+i\varepsilon_1
  \mean{\varsigma_1\varsigma^\dagger_2\varsigma_2 \ud{a}} \Big]\,.
\end{multline}
This leads to the steady state solution:
\begin{equation}
  \label{eq:WedMar21185114CET2012}
  \mean{n_1n_2}=\frac{2}{\Gamma_1+\Gamma_2}\Re \Big[ i\varepsilon_2T_+ \mathbf{w}[11,01]\Big]_1+[1\leftrightarrow 2]\,.
\end{equation}
This solution relies on $\mathbf{w}[11,01]$ which can be expressed in
terms of three lower order correlators:
\begin{multline}
  \mathbf{w}[11,01]= \frac{-1}{M+(-i\omega_2-\Gamma_1-\frac{\Gamma_2}2)\mathbf{1}}\\
  \times \Big\{ -i\varepsilon_2 T_- \mathbf{w}[11,00]-i\varepsilon_1 T_-
  \mathbf{w}[10,01]+i\varepsilon_1 T_+ \mathbf{w}[01,01] \Big\}\,,
\end{multline}
each of which is given by:
\begin{multline}
  \label{eq:ThuMar29195227CEST2012}
  \mathbf{w}[11,00]= \frac{-1}{M-\Gamma_1\mathbf{1}}\\
  \times \Big\{i\varepsilon_1 T_+ \mathbf{w}[01,00]-i\varepsilon_1 T_-
  \mathbf{w}[10,00] \Big\}\,,
\end{multline}
\begin{multline}
  \mathbf{w}[10,01]= \frac{-1}{M+(i\omega_1-i\omega_2-\frac{\Gamma_1+\Gamma_2}2)\mathbf{1}}\\
  \times \Big\{-i\varepsilon_2 T_- \mathbf{w}[10,00]+i\varepsilon_1 T_+ \mathbf{w}[00,01] \Big\}\,,
\end{multline}
and
\begin{multline}
  \label{eq:ThuMar29195317CEST2012}
  \mathbf{w}[01,01]= \frac{-1}{M+(-i\omega_1-i\omega_2-\frac{\Gamma_1+\Gamma_2}2)\mathbf{1}}\\
  \times \Big\{-i\varepsilon_1 T_- \mathbf{w}[00,01]-i\varepsilon_2 T_- \mathbf{w}[01,00] \Big\}\,.
\end{multline}
Finally, we apply a last time the recursive relation
Eq.~(\ref{eq:FriMar30202508CEST2012}) to find the three different
correlators involved in the previous expressions:
\begin{subequations}
  \label{eq:MonJul16004354CEST2012}
  \begin{align}
    &\mathbf{w}[10,00]=
    \frac{-1}{M+(i\omega_1-\frac{\Gamma_1}2)\mathbf{1}} i\varepsilon_1 T_+ \mathbf{v}^\mathrm{ss}\,,\\
    &\mathbf{w}[00,01]=\frac{-1}{M+(-i\omega_2-\frac{\Gamma_2}2)\mathbf{1}} (-i\varepsilon_2
    T_-) \mathbf{v}^\mathrm{ss}\,,\\
    &\mathbf{w}[01,00]=\frac{-1}{M+(-i\omega_1-\frac{\Gamma_1}2)\mathbf{1}} (-i\varepsilon_1
    T_-) \mathbf{v}^\mathrm{ss}\,.
  \end{align}
\end{subequations}

\subsubsection{Sensing method at $\tau>0$}

We now consider the case where the second photon is absorbed by sensor
2 with some delay~$\tau>0$ after a first photon is absorbed by sensor
1. The correlator of interest is $\mean{n_1(0)n_2(\tau)}$, with
equation of motion:
\begin{multline}
  \label{eq:ThuMar29192644CEST2012}
  \partial_\tau \mean{n_1(0)n_2(\tau)}
  =-\Gamma_2\mean{n_1(0)n_2(\tau)}\\+2\Re \Big[ i\varepsilon_2
  \mean{n_1(0)(\varsigma_2 \ud{a})(\tau)} \Big]\,,
\end{multline}
with the initial condition in the steady state
$\mean{n_1(0)n_2(0)}=\mean{n_1n_2}$. This solution relies on
$\mean{n_1(0)(\varsigma_2 \ud{a})(\tau)}$. To compute it, we introduce a
vector analogous to Eq.~(\ref{eq:WedMar21200056CET2012}) but now
consisting of two-time correlators:
\begin{equation}
  \label{eq:WedMar21200056CET2012}
  \mathbf{w}'[11,\mu_2\nu_2](\tau) = \left( \begin{array}{c}
      \mean{n_1(0)(\varsigma_2^{\dagger\mu_2}\varsigma_2^{\nu_2})(\tau)} \\
      \mean{n_1(0)(\varsigma_2^{\dagger\mu_2}\varsigma_2^{\nu_2}a)(\tau)} \\
      \mean{n_1(0)(\varsigma_2^{\dagger\mu_2}\varsigma_2^{\nu_2}\ud{a})(\tau)} \\
      \mean{n_1(0)(\varsigma_2^{\dagger\mu_2}\varsigma_2^{\nu_2}\ud{a}a)(\tau)} \\
      \vdots
    \end{array}
  \right)\,.
\end{equation}
With this definition, $\mean{n_1(0)(\varsigma_2 \ud{a})(\tau)}$ is the
first element of the vector $T_+\mathbf{w}'[11,01](\tau)$. The
$\tau$-equation for $\mathbf{w}'[11,01](\tau)$ reads:
\begin{multline}
  \partial_\tau \mathbf{w}'[11,01](\tau)=\Big[M+(-i\omega_2-\frac{\Gamma_2}2)\mathbf{1}\Big]\mathbf{w}'[11,01](\tau)
  \\-i\varepsilon_2 T_- \mathbf{w}'[11,00](\tau)\,,
\end{multline}
with $\mathbf{w}'[11,00](\tau)=e^{M\tau}\mathbf{w}[11,00]$ with
initial condition $\mathbf{w}'[11,01](0)=\mathbf{w}[11,01]$ in the
steady state. After some algebra, one arrives to the solution:
\begin{multline}
  \mathbf{w}'[11,01](\tau)=e^{\big[M+(-i\omega_2-\frac{\Gamma_2}2)\mathbf{1}\big]\tau}\mathbf{w}[11,01]\\
  -(-i\varepsilon_2)\mathcal{Y}(\tau)\mathbf{w}[11,00]\,,
\end{multline}
in terms of a matrix $\mathcal{Y}(\tau)$ defined elementwise as:
\begin{multline}
  \mathcal{Y}_{ij}(\tau)=\sum_{p,k,l,q}\frac{E_{ip}E^{-1}_{pk}T_-^{kl}E_{lq}E^{-1}_{qj}}{m_p-m_q-i\omega_2-\frac{\Gamma_2}2}\\
\Big\{e^{(m_p-i\omega_2-\frac{\Gamma_2}2)\tau}-e^{m_q\tau} \Big\}\,.
\end{multline}
Substituting this expression into
Eq.~(\ref{eq:ThuMar29192644CEST2012}) and solving it, we obtain:
\begin{multline}
  \label{eq:ThuMar29201228CEST2012}
  \mean{n_1(0)n_2(\tau)}=e^{-\Gamma_2\tau}\mean{n_1n_2}\\
  +2\Re\Big[ i\varepsilon_2
  T_+\mathcal{F}(\tau)\mathbf{w}[11,01]\Big]_1\\
  +2\Re\Big[
  \varepsilon_2^2T_+\mathcal{Z}(\tau)\mathbf{w}[11,00]\Big]_1\,,
\end{multline}
where the matrices $\mathcal{F}(\tau)$ and $\mathcal{Z}(\tau)$ are
those introduced in the previous section, namely,
Eqs.~(\ref{eq:ThuMar29193657CEST2012})
and~(\ref{eq:ThuMar29194113CEST2012}), respectively.


\subsubsection{Equality of the integral and sensing methods}

We complete the proof by showing that the results from the integration
and the sensing methods are the same to leading order in the
coupling~$\varepsilon$.

First, the case $\tau=0$.  The final expression for $\mean{n_1 n_2}$
is obtained by inserting the solutions for the
correlators~(\ref{eq:ThuMar29195227CEST2012}--\ref{eq:ThuMar29195317CEST2012})
into Eq.~(\ref{eq:WedMar21185114CET2012}). This leads to the same
results as Eq.~(\ref{eq:MonMar26124543CEST2012}), with the integrals
appearing precisely in the following order:
\begin{multline}
  \label{eq:WedMar21195020CET2012}
  \mean{n_1 n_2}=\frac{\varepsilon_1^2\varepsilon_2^2}{\Gamma_1\Gamma_2}(2\pi)^2 \\
  \times 2\Re
  \Big\{\mathcal{I}_{(3b)}+\mathcal{I}_{(3a)}+\mathcal{I}_{(2a)}+\mathcal{I}_{(1a)}+\mathcal{I}_{(1b)}+\mathcal{I}_{(2b)}
  \Big\}\\+[1\leftrightarrow 2]
  =\frac{\varepsilon_1^2\varepsilon_2^2}{\Gamma_1\Gamma_2}(2\pi)^2
  S_{\Gamma_1\Gamma_2}^{(2)}(\omega_1;\omega_2)\,.
\end{multline}

Second, the case $\tau>0$. For ease of comparison, we rewrite the term
$\Delta\mathcal{I}$ in term of the vector $\mathbf{w}[11,01]$ as:
\begin{equation}
  \label{eq:ThuMar29200103CEST2012}
  \Delta \mathcal{I}(\tau)= \frac{\Gamma_1\Gamma_2}{(2\pi)^2} \Big[ T_+\mathcal{F}(\tau)\frac{1}{\varepsilon_1^2(-i\varepsilon_2)}\mathbf{w}[11,01]\Big]_1\,.
\end{equation}
It is then clear that this expression is equal, up to a constant
factor, to the second line in the expression for
$\mean{n_1(0)n_2(\tau)}$,
Eq.~(\ref{eq:ThuMar29201228CEST2012}). Similarly, the term $\Delta
\mathcal{I}_{(3\alpha)}(\tau)+\Delta \mathcal{I}_{(3\beta)}(\tau)$ in
Eq.~(\ref{eq:ThuMar29201440CEST2012}) can be rewritten as:
\begin{equation}
  \label{eq:ThuMar29201551CEST2012}
  \Delta \mathcal{I}_{(3\alpha)}(\tau)+\Delta\mathcal{I}_{(3\beta)}(\tau)=\frac{\Gamma_1\Gamma_2}{(2\pi)^2}\Big[T_+
  \mathcal{Z}(\tau)
  \frac{1}{\varepsilon_1^2}\mathbf{w}[11,00]\Big]_1\,,
\end{equation}
and related to the third line in
Eq.~(\ref{eq:ThuMar29201228CEST2012}). All together, we can therefore
conclude that, to leading order in the couplings:
\begin{equation}
  \label{eq:ThuMar29195843CEST2012}
  \mean{n_1(0)n_2(\tau)}=\frac{\varepsilon_1^2\varepsilon_2^2}{\Gamma_1\Gamma_2}(2\pi)^2
  S_{\Gamma_1\Gamma_2}^{(2)}(\omega_1;\omega_2,\tau)\,.
\end{equation}

\subsection{Final remarks}

This proof can be generalised to $N$-photon correlations and/or for
finite $T_1$-time dynamics (instead of a steady state) by repeating
these procedures linearly in the number of sensors and
integrals. There is no conceptual difference brought by the higher
number of variables, but notations become heavy and for the sake of
clarity, we have illustrated the proof in the simplest, as well as
most relevant cases, of $N=1$ and~$2$. Also, nothing in the proof
relies on the choice of sensors as two-level systems, which has been
made for convenience. As we always examine crossed correlations
between them, they could also be, e.g., harmonic oscillators, and
provide identical results.

Together with
Eqs.~(\ref{eq:WedMar21185114CET2012}--\ref{eq:MonJul16004354CEST2012}),
Eq.~(\ref{eq:ThuMar29201228CEST2012}) provides a semi-analytical
result that can be used directly for computations. Although the
Hilbert space is not enlarged when using these formulas, they are
however awkward to use and set up.  Also, the growth in the number of
correlators has the same power dependence on the maximum number of
excitations allowed in the system than when including the sensors
explicitly (it is linear in the Jaynes--Cummings model).  The number
of correlators increases like $4^N$ when including $N$ sensors.
Benchmarks for $N=2$ show that the many matrix operations (inversions
and multiplications) involved to evaluate the formulas are more costly
than solving linear equations as required when including explicitly
the sensors.  Although this is for a larger set of correlators in the
latter case, optimisations such as LU decomposition make sensors a
more efficient as well as a conceptually simpler approach. If using
the semi-analytical formulas turns out to be more effective in a
particular context or for larger $N$, similar results can be derived
for $\langle n_1(0)n_2(\tau_1)\dots n_N(\tau_{N-1})\rangle$ by
generalizing Eq.~(\ref{eq:FriMar30202508CEST2012}) with $N$ sensors to
obtain $\mathbf{w}[\mu_1\nu_1,\dots,\mu_N\nu_N]$ recursively.

Finally, the case of $N$ identical sensors reproduces exactly the
$N$-photon correlations, $g^{(N)}$, from a single harmonic sensor
(full correlations of the output of a single filter). This can be
shown by comparing the presented derivation with $N$ two-level sensors
with one where the system is coupled to a single bosonic sensor with
associated $\mathbf{w}$ vectors of the type $\mathbf{w}[n,m]$ (where
$n$, $m=0,\dots N$). The results are also seen to be identical to
those obtained by substituting $\omega_1=\omega_2=\omega$ and
$\Gamma_1=\Gamma_2=\Gamma$ in the formula for
$g_{\Gamma_1\Gamma_2}^{(2)}(\omega_1;\omega_2,\tau)$.

\section{Further application to the Jaynes--Cummings model}

\begin{figure*}[t] 
  \centering
  \includegraphics[width=.97\linewidth]{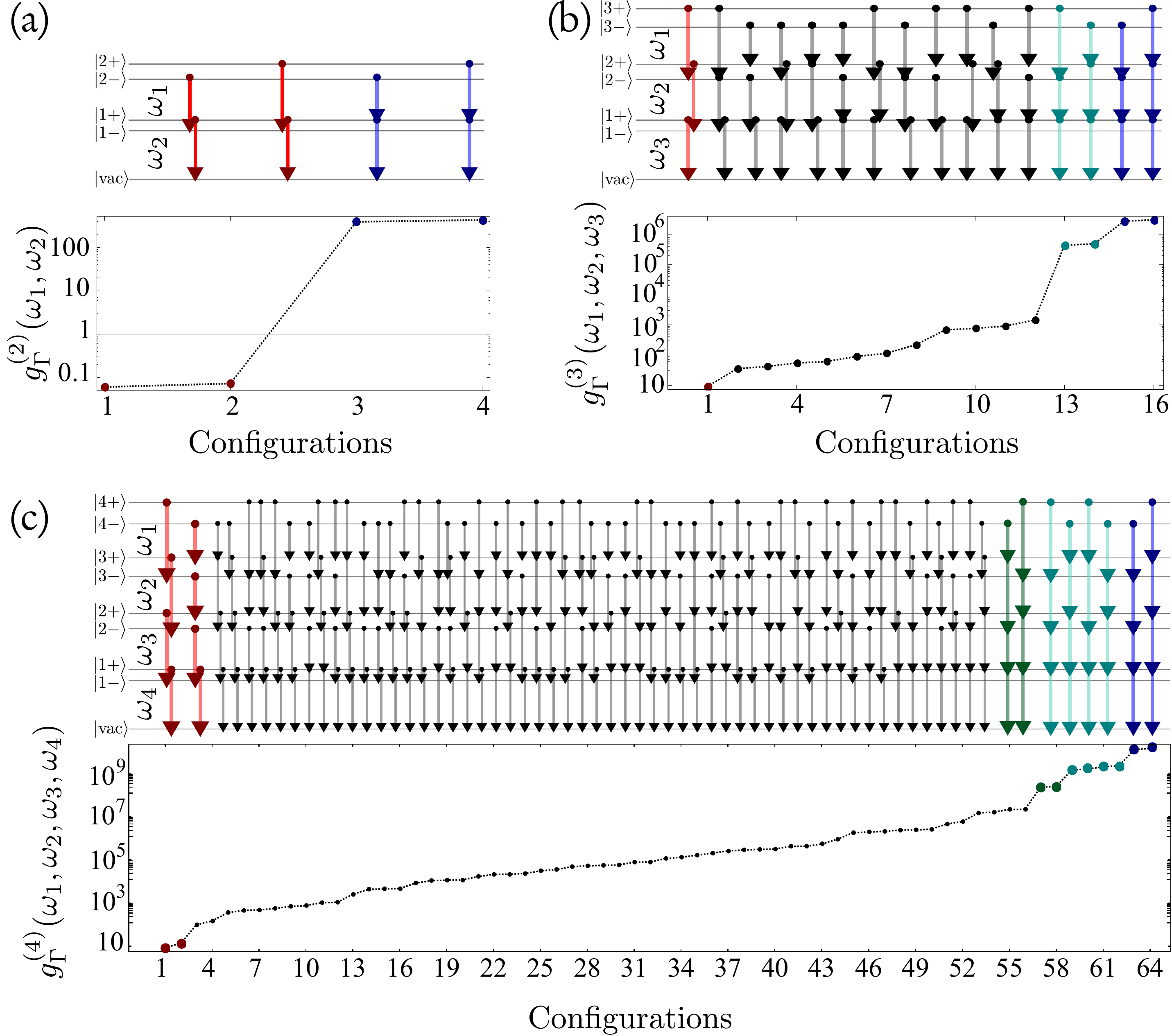}
  \caption{(Color online) (a) Two-, (b) three- and (c) four-photon
    correlations in the Jaynes--Cummings model with detection at the
    frequencies corresponding to transitions in the ladder. For
    visibility, whenever the arrows would overlap, a small shift has
    been introduced to assist the eye in tracking the starting and
    ending points. By probing transitions in a cascade or, on the
    contrary, from different de-excitation routes, the correlations
    between the detected photons vary over 3, 5 and 10 orders of
    magnitudes at the two, three and four-sensor level,
    respectively. Note that the vertical axes are in log-scale, so
    even if the variations appear moderate, they are locally
    important. Only transitions detecting the upper polariton
    $\ket{1+}\rightarrow\ket{\mathrm{vac}}$ in the first rung are
    shown, those detecting the lower polariton
    $\ket{1-}\rightarrow\ket{\mathrm{vac}}$ are reconstructed by
    swapping upper and lower polaritons in all rungs.  Parameters:
    $\gamma_a=\gamma_\sigma=0.001g$, $\Gamma=\gamma_2=0.003g$ in the
    limit of vanishing incoherent pumping of the emitter,
    $P_\sigma\rightarrow0$.}
  \label{fig:FriJul13220009CEST2012}
\end{figure*}

The sensing method was illustrated in the text up to three frequencies
and for various time delays. Here we provide a supplemental example up
to four-photon correlations.

Such high-order correlations are not intuitive to visualise in their
most general representation, given that they convey more information
and of a much deeper character than single-photon observables.
Photons are emitted at all energies and some correlations for
particular energies other than $\pm R_n^\pm$ are suppressed or, on the
contrary, enhanced, meaning that more complicated processes than
simple relaxation take place. We reserve to future works the
presentation of how such new processes of emission can be identified
in the study of frequency resolved correlations, already at the
two-photon level, and how these may find new applications to optimise
quantum emitters. Here, to keep the discussion succinct, we will focus
on the most important processes only, where $N$ photons are detected
at precisely the Jaynes--Cummings transitions, that is, we disregard
the correlations where one or more photons have an energy which does
not correspond to a transition in the ladder.

In Fig.~\ref{fig:FriJul13220009CEST2012}, we compare two, three and
four-photon correlations of photons with energies corresponding to the
possible placements of detectors over all possible transitions.  There
are $2^{2N-1}$ configurations, half of them being symmetric with the
other half by the interchange of upper and lower polaritons in all
rungs. We need only consider, therefore, $2^{2N-2}$ cases, which are
displayed in Fig.~\ref{fig:FriJul13220009CEST2012} representing only
the half which detects the upper polariton in the first rung.  For
instance the leftmost case in panel (c) corresponds to setting four
detectors at the energies $\omega_1$, \dots, $\omega_4$ probing the
transitions $\ket{4+}\rightarrow\ket{\mathrm{3-}}$,
$\ket{3+}\rightarrow\ket{\mathrm{2-}}$,
$\ket{2+}\rightarrow\ket{\mathrm{1-}}$ and
$\ket{1+}\rightarrow\ket{\mathrm{vac}}$. As in this sequence of
detection, polaritons have to swap branch in all rungs, the emission
is unlikely and the corresponding coincidence is strongly suppressed.

As discussed in the main text, finite $\tau$ are important since
correlations may be maximised at nonzero time-delays, when the
dynamics of relaxation synchronises with detection. It is however
difficult to find the optimising values for $N-1$ independent degrees
of freedom when measuring $N$th order correlations. We show here that
the simplest approximation to fix all delays at zero already leads to
useful results which contain the gist of the dynamics. An absolute
value of photon correlations has little meaning in itself. It is when
compared to other correlations in alternative configurations that a
physical meaning can be identified and
quantified. Figure~\ref{fig:FriJul13220009CEST2012} shows how, even at
equal times, the detection of photons at energies that correspond to a
cascade of the Jaynes--Cummings ladder results in giant
bunching. These are the points on the right of each panel. On the
opposite, as previously described, when the detectors are arranged to
click in the sequence that least correspond to a cascade, that is,
alternating the type of polariton each time the system goes one rung
down, a corresponding giant suppression is obtained. This is an actual
antibunching in the case of two-photon detection (a), while with a
higher number of photons, the values obtained are larger than one but,
again, when compared to the relative values of other transitions,
reveal a giant suppression of correlations of over five and ten orders
of magnitudes in three and four-photon counting, respectively.

Another remarkable behaviour of these figures is the emergence of a
classical behaviour with the increasing number of detected photons,
powered by combinatorial growth. While correlations are markedly
distinct and varying abruptly in the extreme, low-entropy situations
(on both sides of the horizontal axes), the large number of
intermediate configurations smoothes out the quantized character and
yields a gradual and milder variation as one quantum in the chain of
detections is shifted from its precise expected value. The larger the
number of photons, the faster is this transition from a discrete,
staircase behaviour to a smooth continuous one. These are the
transitions shown in black (also with smaller arrows in~(c)). These
results also reveal that proper sequencing of the detection allows to
isolate and magnify its quantum character, even when dealing with a
large number of photons. This only hints at the rich physics
unravelled by $N$-photon correlations and at the applications they
could bring about.

\bibliography{Sci,books}